\documentclass[prl,twocolumn,showpacs, superscriptaddress]{revtex4-1}
\usepackage{placeins}
\usepackage{amsfonts}
\usepackage{amssymb}
\usepackage{amsmath, mathtools}
\usepackage{graphicx}
\usepackage{datetime}
\usepackage{color}
\usepackage{enumerate}
\usepackage{pifont,array}
\usepackage[normalem]{ulem}

\newcommand{\ket}[1]{|#1\rangle}
\newcommand{\bra}[1]{\langle#1|}
\newcommand{\tr}[1]{\mathrm{tr}(#1)}
\newcommand{\expectn}[1]{\langle#1\rangle}
\renewcommand{\mid}{\,{|}\,}

\begin{document}

\title{Residual and Destroyed Accessible Information after Measurements}

\author{Rui Han}
\affiliation{Max Planck Institute for the Science of Light, 91058 Erlangen, Germany}
\affiliation{Institut f\"ur Optik, Information und Photonik, Universit\"at Erlangen-N\"urnberg, 91058 Erlangen, Germany}
\email{han.rui@quantumlah.org}

\author{Gerd Leuchs}
\affiliation{Max Planck Institute for the Science of Light, 91058 Erlangen, Germany}
\affiliation{Institut f\"ur Optik, Information und Photonik, Universit\"at Erlangen-N\"urnberg, 91058 Erlangen, Germany}
\affiliation{University of Ottawa, Ottawa ON K1N 6N5, Canada}

\author{Markus Grassl}
\affiliation{Max Planck Institute for the Science of Light, 91058 Erlangen, Germany}

\begin{abstract}
  When quantum states are used to send classical information, the
  receiver performs a measurement on the signal states. The amount of
  information extracted is often not optimal due to the receiver's
  measurement scheme and experimental apparatus. For quantum
  non-demolition measurements, there is potentially some residual
  information in the post-measurement state, while part of the
  information has been extracted and the rest is destroyed. Here, we
  propose a framework to characterize a quantum measurement by how
  much information it extracts and destroys, and how much information
  it leaves in the residual post-measurement state. The concept is
  illustrated for several receivers discriminating coherent states.
\end{abstract}

\pacs{03.67.Hk, 
  42.79.Sz, 
  32.80.-t 
}

\maketitle

\textit{Introduction.}--- Quantum measurements are often associated
with the expectation value of an observable, which corresponds to a
physical quantity, such as the average energy of a system or the mean
photon number.  For this, one has to make repeated measurements on
identically prepared copies of a quantum state (ensemble average).  In
the context of quantum information, on the other hand, one usually
considers one-shot measurements.  The result of the measurement is
described as one out of $M$ possible outcomes, and the measurement
provides some classical information about the quantum state.  Quantum
state discrimination is a special case of this scenario.  The receiver
who performs the measurement knows that the state he receives is from
a set of given quantum states with fixed prior probabilities, and he
only needs to identify which state it is.

In the following scenario, referred to as classical-quantum (cq)
communication~\cite{NielsenChuang:00}, Alice encodes her classical
information using a given set of (orthogonal or non-orthogonal)
quantum states and sends a particular signal state to Bob.  Bob
constructs a set of measurements on the signal he receives to decode
the information from his measurement outcomes. In order to make the
communication channel between Alice and Bob as efficient or secure as
possible, Bob should not count on having identical copies of the same
quantum state, but instead, make the most use of every copy he
receives.

In order to characterize the communication channel between Alice and
Bob, one often uses the average error probability or the mutual
information. For a certain class of pure quantum signals, the average
probability for Bob making an error when decoding Alice's signal is
minimized by the so-called square-root measurement or by the Helstrom
measurement~\cite{Helstrom:76, Ban1997, Eldar:2001}.  From a
communication perspective, the mutual information, quantifying how
much information is transmitted between Alice and Bob, is the more
relevant figure of merit~\cite{wilde_2013}.  The two concepts are not
equivalent as, for example, minimizing the error probability does not
necessarily result in maximal mutual information.

Bob extracts information about Alice's state through his measurement
outcome, and the amount of Bob's information is upper bounded by the
so-called accessible information.  In general, it is very often not
possible for Bob to implement an optimal measurement attaining the
upper bound.  When Bob performs a von Neumann measurement given by
rank-one projections, the state after the measurement carries no
additional information as it only depends on the measurement outcome,
but no longer on the initial state.  Hence, the information that has
not been extracted by Bob is fully destroyed.  On the other hand, if
Bob performs a generalized quantum measurement -- typically referred
to as positive-operator valued measure (POVM) or probability operator
measure (POM) -- with operators of rank larger than one, the
post-measurement state could still contain some information about the
input state.  That residual information can be extracted through a
subsequent measurement to increase Bob's total information
gain~\cite{PhysRevLett.77.2154, PhysRevLett.96.020408,
  PhysRevA.73.012113, Nagali2012}.  How much information is extracted
by the measurement depends only on the POVM element.  The amount of
residual information left in the post-measurement state, however,
depends on the very operators used to implement the POVM measurement.

In the full realm of quantum mechanics, very often, the error
probability or the gain of knowledge have been used to quantify the
effectiveness of measurements, and fidelity measures have been used to
quantify the disturbance of measurements on a quantum
state~\cite{0305-4470-34-35-303, Barnum:2002, PhysRevLett.96.020408,
  PhysRevA.73.012113, Nagali2012}. In the present work, however, we
fully characterize a quantum measurement using mutual information as
the figure of merit, more specifically the amount of information that
is extracted, how much information is destroyed, and how much is
left-over in the post-measurement state. We illustrate, with practical
examples, the power of this approach by looking at four different
measurement schemes for binary coherent-state discrimination.

\textit{Quantum measurements}.--- In the general scenario of
classical-quantum communication, Alice encodes the information using
an ensemble $\mathcal{E}$ of $N$ signal states $\{\rho_j\colon
j=1,2,\dots,N\}$, which can be mixed or pure, with prior probability
distribution $\{\eta_j\colon j=1,2,\dots,N\}$.  Bob, in order to
identify the state he receives from Alice, can perform any von Neumann
or generalized measurement on the state.  The measurement can be
either direct, i.e., measuring the state itself, or indirect by
entangling the state to an ancilla system first and then measuring the
ancilla \cite{Neu40}.  We describe Bob's
measurement by an $M$-element POVM on the signal state:
$\Pi\equiv\{\Pi_k\colon k=1,2,\dots,M\}$, $\sum_{k=1}^M
\Pi_k=\openone$.  When Bob wants to establish a one-to-one
correspondence between the measurement outcomes and the full set of
signal states, one clearly needs $M\ge N$. For $M>N$, the most simple
scheme is obtained by grouping the POVM elements, for example, Bob
could associate the measurement outcomes of $\Pi_1,\dots,\Pi_{k_1}$
with the state $\rho_1$.

The initial knowledge of the signal state can be represented by the
statistical operator $\rho=\sum_{j=1}^N\eta_j\rho_j$, where
$\tr{\rho}=1$.  The joint probability that the state $\rho_j$ is sent
and that the measurement outcome is $\Pi_k$, is given by $P(\rho_j,
\Pi_k)=\eta_j\tr{\Pi_k\rho_j}$. The marginal over the label $j$ of the
input states,
\begin{equation}
  P_{\Pi_k}=\sum_{j=1}^N \eta_j\tr{\Pi_k\rho_j}=\tr{\Pi_k\rho}\,,
\end{equation}
gives the total probability of having measurement outcome $\Pi_k$, and
the marginal over the label $k$ of the measurement outcomes,
$\sum_{k=1}^M P(\rho_j,\Pi_k)=\eta_j$, is just the prior probability of the
state $\rho_j$.  The mutual information,
\begin{equation}\label{eq:mutual_information}
I(\mathcal{E}\,{:}\,\Pi)=H(\mathcal{E})-\sum_{k=1}^MP_{\Pi_k}H(\mathcal{E}|\Pi_k)
\end{equation}
quantifies how much information is shared between Alice and Bob
through Bob's POVM measurement $\Pi$.  The Shannon entropy of Alice's
signal is given by $H(\mathcal{E})=-\sum_{j=1}^N \eta_j {\log_2}
\eta_j$. The conditional entropy $H(\mathcal{E}\mid\Pi_k)$ quantifies
Bob's remaining ignorance about the signal state given the measurement
outcome $\Pi_k$.  The accessible information of the ensemble
$\mathcal{E}$ is defined as the maximal mutual information attainable
over all possible POVMs,
\begin{equation}\label{eq:accessible_information}
  I_\mathrm{acc}(\mathcal{E})=H(\mathcal{E})-\min_{\mathrm{all\;}\Pi}\sum_{k=1}^M P_{\Pi_k}H(\mathcal{E}\mid\Pi_k)\,.
\end{equation}
The accessible information and the set of optimal measurements is
known in closed form only for very few special cases, namely, for a
communication channel with pure binary states or with real-symmetric
trine states~\cite{holevo:73b, Levitin1995, Sasaki1999}.  In general,
the accessible information is usually obtained using numerical
optimization methods~\cite{Peres1991, Osaki:1998, Shor2002,
  PhysRevA.71.054303}.  Holevo's theorem provides an upper bound on
the accessible information in terms of the so-called Holevo quantity.
Although the Holevo bound is asymptotically achievable when collective
measurements on a large number of signals are allowed, it is very
often not tight when only single-copy measurements are allowed
\cite{Holevo:73a, holevo:73b, PhysRevLett.73.3047, Hausladen1996}.

When Bob's POVM does not extract all possible information, he could,
at least in principle, perform a subsequent measurement on the
post-measurement state to proceed further.
Each POVM element corresponds to a general quantum operation with
Kraus operator $A_k$, where $\Pi_k=A_k^\dagger A_k$~\footnote{For
  simplicity, we restrict ourselves to the construction of the POVM
  with single Kraus operators here, but the formalism directly
  generalizes to the case when $\Pi_k=\sum_\ell A_{k\ell}^\dagger
  A_{k\ell}$.}. When the measurement outcome for $\Pi_k$ is obtained,
the normalized post-measurement state $\rho_j^{(k)}$ corresponding to Alice's
state $\rho_j$ and the new prior probabilities are~\cite{KRAUS1971311}
\begin{equation}\label{eq:PostmMasurementStates}
  \rho_j^{(k)}=\frac{A_k\rho_jA_k^\dagger}{\tr{\Pi_k\rho_j}}\;\;\text{and}\;\;\eta_j^{(k)}=\frac{\eta_j\tr{\Pi_k\rho_j}}{\tr{\Pi_k\rho}}\,.
\end{equation}
They form the ensemble of post-measurement states $\mathcal{E}^{(k)}$,
conditioned on a particular measurement outcome $\Pi_k$.  Note that we
can also express the conditional Shannon entropy
$H(\mathcal{E}\mid\Pi_k)$ in Eqs.~\eqref{eq:mutual_information} and
\eqref{eq:accessible_information} as the Shannon entropy
$H(\mathcal{E}^{(k)})$.  To discriminate the post-measurement states,
Bob then can perform any subsequent POVM.  For an optimal subsequent
measurement on $\mathcal{E}^{(k)}$, the remaining ignorance about the
ensemble $\mathcal{E}$ is reduced to
$H(\mathcal{E}^{(k)})-I_\mathrm{acc}(\mathcal{E}^{(k)})$.  Then the
maximal mutual information between the ensemble $\mathcal{E}$ of
signal states and the outcomes of optimal subsequent measurements is
given by
\begin{equation}\label{eq:accessible_info2}
  I'_{\text{max}}(\mathcal{E},\Pi)=
H(\mathcal{E})- \sum_{k=1}^M P_{\Pi_k}\left[H(\mathcal{E}^{(k)})-I_\mathrm{acc}(\mathcal{E}^{(k)})\right],
\end{equation}
which only depends on $\mathcal{E}$ and Bob's first measurement $\Pi$.
Note that Bob's final message solely depends on the outcome of the
subsequent measurement, because the result of the first measurement is
incorporated in the updated new prior probabilities $\{\eta_j^{(k)}\}$
for the discrimination of $\{\rho_j^{(k)}\}$.  Therefore,
$I'_{\text{max}}(\mathcal{E},\Pi)$ is never smaller than the mutual
information $I(\mathcal{E}\,{:}\,\Pi)$ of the first measurement, i.e.,
$I'_{\text{max}}(\mathcal{E},\Pi)\ge I(\mathcal{E}\,{:}\,\Pi)$.
Equality holds if and only if the post-measurement states are
independent of the input state, i.e., when
$I_{\text{acc}}(\mathcal{E}^{(k)})=0$ for all $k$.

\textit{Information-theoretic characterization}.---
The efficiency of a measurement $\Pi$ in attaining information
can be quantified by the fraction of information extracted, defined as
\begin{equation}
  \bar{E}\equiv\frac{I(\mathcal{E}\,{:}\,\Pi)}{I_\mathrm{acc}(\mathcal{E})}\,.
\end{equation}
The amount of extracted information is normalized by the total
accessible information $I_\mathrm{acc}(\mathcal{E})$ such that
$0\leq\bar{E}\leq1$. When information is not fully extracted by the
measurement, i.e., $\bar{E}<1$, part of the information can still be
preserved in the post-measurement state and hence might be accessible via suitable
subsequent measurements.
Thus, we define the fraction of residual information that can potentially
be extracted via subsequent measurements as
\begin{equation}
  \bar{R}\equiv\frac{I'_{\text{max}}(\mathcal{E},\Pi)- I(\mathcal{E}\,{:}\,\Pi)}{I_\mathrm{acc}(\mathcal{E})}\,.
\end{equation}
The residual information is bounded by $0\leq\bar{R}\leq 1-\bar{E}$.

The maximal mutual information $I'_{\text{max}}(\mathcal{E},\Pi)$
achievable by any multi-step protocol with a particular first
measurement $\Pi$ performed by the receiver cannot exceed the
accessible information $I_{\text{acc}}(\mathcal{E})$ of the original
signal states, thus, $I'_{\text{max}}(\mathcal{E},\Pi)\le
I_{\text{acc}}(\mathcal{E})$.  The first measurement does not destroy
any information if and only if $I'_{\text{max}}(\mathcal{E},\Pi)=
I_{\text{acc}}(\mathcal{E})$.  Hence, we define the fraction of
information destroyed as
\begin{equation}
  \bar{D}\equiv\frac{I_{\text{acc}}(\mathcal{E})- I'_{\text{max}}(\mathcal{E},\Pi)}{I_{\text{acc}}(\mathcal{E})}\,,
\end{equation}
which quantifies the reduction of accessible information due to the
measurement $\Pi$.  Combing the three parts---the fraction of
extracted information $\bar{E}$, residual information $\bar{R}$, and
destroyed information $\bar{D}$, respectively---we have conservation
of total accessible information,
\begin{equation}
  \bar{E}+\bar{R}+\bar{D}=1\,.
\end{equation}
Here, we choose to use the accessible information, which is computed
in \eqref{eq:accessible_information} via an optimization over all
possible measurements, as the conserved quantity and to normalize
other quantities by. When the accessible information is not known, we
can replace $I_{\text{acc}}(\mathcal{E})$ by the mutual information for
a sub-optimal measurement $\widetilde{\Pi}$ for the task at hand (such
as the Helstrom measurement). Then the corresponding quantities are
 defined in relation to $\widetilde{\Pi}$.

\textit{Examples}.---
In the following, we illustrate the significance of characterizing a
quantum measurement by $\bar{E}$, $\bar{R}$, and $\bar{D}$ in the
scenario of binary coherent-state discrimination, an important example
for classical-quantum optical communication.  For the discrimination
of binary coherent states $\{\ket{\alpha},\ket{{-\alpha}}\}$ with prior
probabilities $\{\eta_1,\eta_2\}$, the well-known Helstrom
measurement is not only the measurement that minimizes the average
error probability but also the measurement that maximizes the mutual
information~\cite{Levitin1995}.  Therefore, for such a given set of signals, the
amount of accessible information $I_{\text{acc}}(\mathcal{E})$ is known,
and it is less than unity owing to the intrinsic non-orthogonality
among coherent states.  Although the mathematical construction of the
Helstrom measurement has been known for decades, it has not yet been
experimentally realized due to limitations in both the experimental
apparatus and receiver strategies \footnote{The Helstrom measurement
  would require the receiver to project the state onto a superposition
  of the two coherent signal states.}.

For discrimination schemes that use a two-element POVM
$\{\Pi_1,\Pi_2\}$, including the Helstrom measurement, the measurement
outcomes of $\Pi_1$ and $\Pi_2$ are associated with the signal states
$\ket{\alpha}$ and $\ket{{-}\alpha}$, respectively. The success
probabilities for identifying the states are
$\{p_1=\expectn{\alpha|\Pi_1|\alpha},
\,p_2=\expectn{{-\alpha}|\Pi_2|{-\alpha}}\}$, and the error
probabilities are $\{r_1=\expectn{\alpha|\Pi_2|\alpha},
\,r_2=\expectn{{-\alpha}|\Pi_1|{-\alpha}}\}$. The measurement outcome
for $\Pi_1$ occurs with probability
$P_{\Pi_1}=\eta_1p_1+\eta_2r_2=\eta_1(1-r_1)+\eta_2r_2$, and for
$\Pi_2$ with probability $P_{\Pi_2}=1-P_{\Pi_1}$.  The mutual
information extracted by this measurement is
\begin{equation}\label{eq:mutual_info}
  I(\mathcal{E}\,{:}\,\Pi)=H(\eta_1)-P_{\Pi_1}H\!\left(\frac{\eta_2r_2}{P_{\Pi_1}}\right)\!-P_{\Pi_2}H\!\left(\frac{\eta_1r_1}{P_{\Pi_2}}\right)\!,
\end{equation}
where $H(p)=-p{\log_2}(p)-(1-p){\log_2}(1-p)$ gives the binary Shannon entropy
for a binary random variable with distribution $\{p,1-p\}$.  Thus,
evaluating \eqref{eq:mutual_info} at the minimum error probabilities
$r_{1}$ and $r_2$ given by the Helstrom measurement yields the
accessible information
\begin{equation}\label{eq:acc_info}
  I_{\text{acc}}(\mathcal{E})=I(\mathcal{E}\,{:}\,\Pi)\Big|_{r_{1,2}=\frac{1}{2}\!\left(\!1-\frac{1-2\eta_{2,1}|\expectn{\alpha|{-\alpha}}|^2}{\sqrt{1-4 \eta_1 \eta_2|\expectn{\alpha|{-\alpha}}|^2}}\!\right)}\!.
\end{equation}
The maximal mutual information $I'_{\text{max}}(\mathcal{E},\Pi)$ for
the subsequent measurement can then be evaluated from a Helstrom
measurement on the post-measurement states.

\begin{figure}[t]
  \centerline{\setlength{\unitlength}{1pt}
    \begin{picture}(240,225)(0,5)
      \put(0,0){\includegraphics[width=240pt]{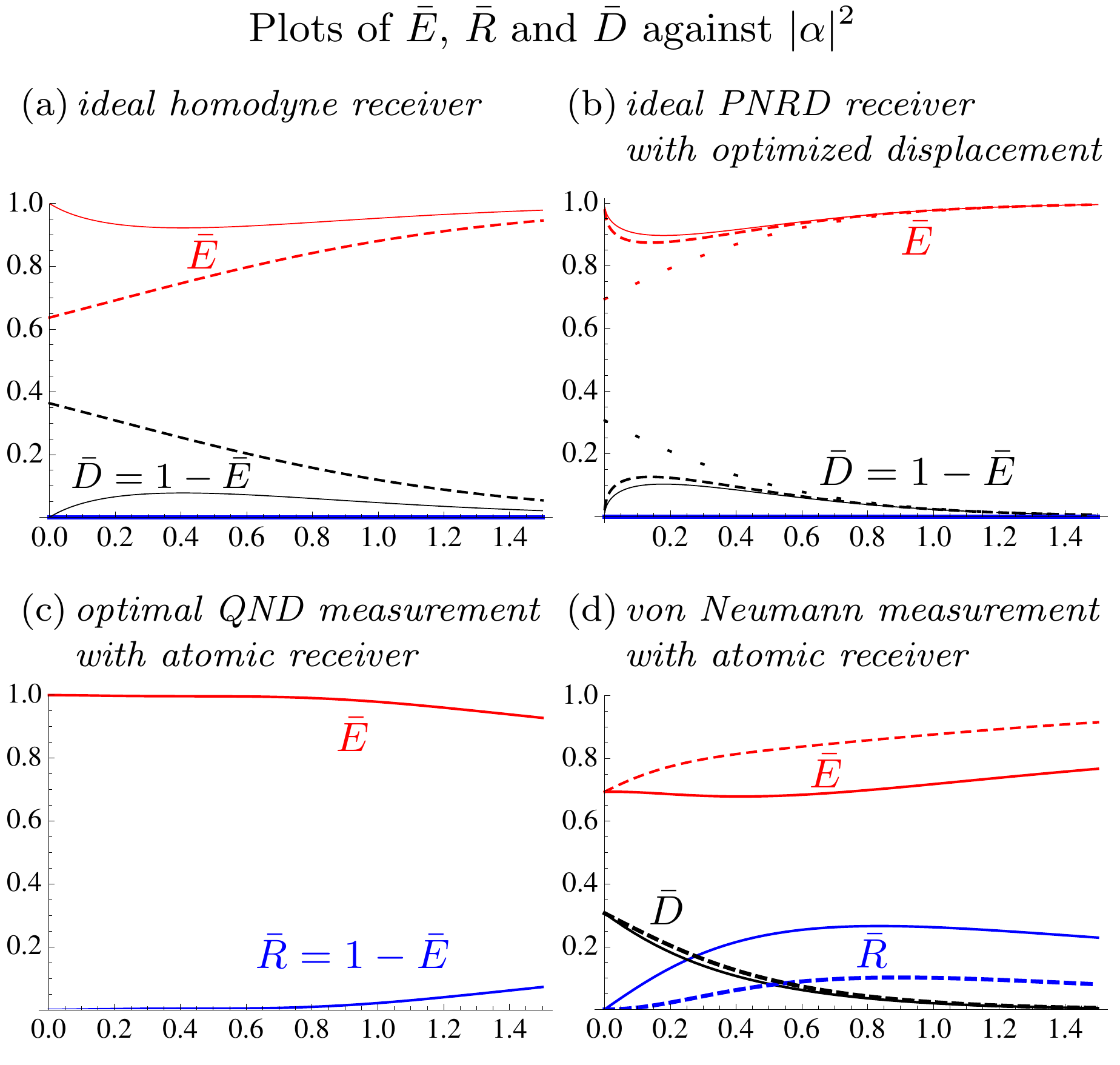}}
   \end{picture}}
  \caption{The fraction of information extracted
    $\bar{E}$ (red), the fraction of information destroyed $\bar{D}$
    (black), and the fraction of residual information $\bar{R}$ (blue)
    are plotted against the mean photon number $|\alpha|^2$ for the
    discrimination of binary coherent states
    $\{\ket{\alpha},\ket{{-\alpha}}\}$ with equal priors.  The
    hard-decision scheme for the homodyne receiver (a) and PNRD
    receiver (b) is illustrated by the dashed curves, and the
    soft-decision scheme is illustrated by the solid curves. In (b),
    the dotted curves are plotted for the Kennedy receiver with
    displacement operation $D(\beta=\alpha)$.  The schemes of the
    non-destructive implementation using an atomic receiver described
    in Ref.~\cite{Han:2017b} are illustrated by (c) and (d).  The data
    for the optimal scheme where the average error probability is
    minimized are plotted in (c), and (d) shows the von Neumann
    measurement scheme that unambiguously discriminates the signal
    state $\ket{\alpha}$ with a single measurement (solid curves) and two sequential measurements (dashed curves).}\label{fig:1}
\end{figure}

The minimum error probability attainable by all Gaussian field
detectors is achieved by the perfect homodyne
receiver~\cite{PhysRevA.78.022320}.  In the hard-decision scheme, a binary
decision is made upon the sign of the measured quadrature. The
corresponding POVM elements are $\Pi_1=\int_0^\infty
dx\,\ket{x}\bra{x}$ and $\Pi_2=\int_{-\infty}^0 dx\, \ket{x}\bra{x}$,
where $\ket{x}$ denotes the state with quadrature value $x$.  The
probability that a coherent state $\ket{\alpha}$ has a measured
quadrature $x$ using a balanced homodyne detector is
$|\expectn{\alpha|x}|^2=\sqrt{2/\pi}e^{-2(x-|\alpha|)^2}$.  The
probabilities of error $r_1=r_2=[1-\mathrm{erf}(\sqrt{2}|\alpha|)]/2$
are identical for the two signal states, and (\ref{eq:mutual_info})
gives the mutual information for this two-element POVM.  The maximum
mutual information attainable using a homodyne receiver is, however,
only achieved by the soft-decision scheme regarding each measured
quadrature value as a measurement outcome of the projector
$\Pi_x=\ket{x}\bra{x}$ in the continuous space of quadratures.  The
hard- and soft-decision schemes yield the same average error
probability, but the amount of extracted information with the
soft-decision scheme is significantly larger; see Fig.~\ref{fig:1}(a). Since there is no
access to the post-measurement state for such a homodyne receiver, the
information that fails to be extracted is completely destroyed, i.e.,
$\bar{D}=1-\bar{E}$.

Another class of popular schemes uses
photon-number-resolving-detection (PNRD) receivers that discriminate
coherent states by their photon statistics, which is a non-Gaussian
property of the field~\cite{PhysRevLett.104.100505,
  PhysRevLett.106.250503, Becerra:14, PhysRevLett.117.200501}. A displacement operator
$D(\beta)$ displaces the signal states
$\{\ket{\alpha},\ket{{-\alpha}}\}$ to
$\{\ket{\alpha+\beta},\ket{\beta-\alpha}\}$ before the signal is sent
to the photon-number-resolving detectors.  In the hard-decision
scheme, a binary decision is made upon whether photons are detected or
not.  In the Fock basis, the POVM elements corresponding to the two
signal states are $\Pi_1=\sum_{j=1}^\infty\ket{j}\bra{j}$ and
$\Pi_2=\ket{0}\bra{0}$.  The probabilities for wrongly identifying the
states $\ket{\alpha+\beta}$ and $\ket{\beta-\alpha}$ are
$r_1=|\expectn{0|\alpha+\beta}|^2=e^{-|\alpha+\beta|^2/2}$ and
$r_2=1-|\expectn{0|\beta-\alpha}|^2=1-e^{-|\beta-\alpha|^2/2}$,
respectively. The soft-decision scheme for the PNRD receiver fully
takes into account each specific measurement outcome of the POVM given
by projections onto all elements of the Fock basis:
$\{\Pi_j=\ket{j}\bra{j}\colon j=0,1,2,\dots\}$.  The difference in the
fraction of information extracted $\bar{E}$ between the hard- and
soft-decision scheme is large for a very weak light field, and
becomes smaller as the field amplitude $|\alpha|$ increases; see
Fig.~\ref{fig:1}(b).  Similar to the homodyne receiver, the residual
light field is completely destroyed by the detector, and hence
$\bar{D}=1-\bar{E}$.

The Neumark dilation theorem~\cite{Neu40} enables the implementation
of any two-element POVM by entangling the signal to a qubit
ancilla and measuring the ancilla system. This process is described by
\begin{alignat}{5}
  U\ket{\alpha}\ket{\mathrm{i}}&{}=\sqrt{p_1}\ket{\varphi_1}\ket{1}+\sqrt{r_1}\ket{\phi_1}\ket{2}\,,\nonumber\\
  U\ket{{-\alpha}}\ket{\mathrm{i}}&{}=\sqrt{r_2}\ket{\varphi_2}\ket{1}+\sqrt{p_2}\ket{\phi_2}\ket{2}\,,\label{unitary}
\end{alignat}
where $U$ is the unitary entangling operation and $\ket{\mathrm{i}}$ denotes
the initial state of the ancilla qubit in the Hilbert space spanned by
the orthogonal basis $\{\ket{1},\ket{2}\}$.  The ancilla state is
measured using projections $\Pi_1=\ket{1}\bra{1}$ and
$\Pi_2=\ket{2}\bra{2}$.  The Helstrom measurement can be effectively
implemented by optimizing the unitary operator $U$~\cite{Han:2017a}.
In practice, however, the set of implementable POVMs is limited by the
choice of the physical ancilla and the available unitary
operations/couplings between the field and the ancilla.  Since the
measurement is only on the ancilla, thus, non-destructive on the light
state, additional information could be extracted by discriminating the
post-measurement states $\{\ket{\varphi_1},\ket{\varphi_2}\}$ when the
measurement outcome is $\Pi_1$, or discriminating the post-measurement
states $\{\ket{\phi_1},\ket{\phi_2}\}$ when the outcome is $\Pi_2$.

Ref.~\cite{Han:2017b} investigated the implementation of such
non-destructive measurements with the Jaynes-Cummings interaction
between the light signal and a two-level atomic ancilla.  Effectively,
the atom serves as a receiver where the information carried by the
coherent state is transferred to and then measured.  
The optimal minimum-error discrimination strategy corresponds to
initially preparing the atom in its ground state $\ket{\mathrm{g}}$ and, after
its interaction with the light field, projecting it onto the equal
superposition states of the ground and excited states
$\{\ket{1}=(\ket{\mathrm{g}}-i\ket{\mathrm{e}})/\sqrt{2},\,\ket{2}=(\ket{\mathrm{g}}+i\ket{\mathrm{e}})/\sqrt{2}\}$.
The minimum error probability for this scheme can be extremely close
to the Helstrom bound for weak coherent signals, i.e., $\bar{E}$ is
very close to unity when $|\alpha|^2$ is small. Moreover, this scheme
also fully preserves in the post-measurement states any information
that has not yet been extracted,
i.e., $I'_{\text{max}}(\mathcal{E},\Pi)=I_{\text{acc}}(\mathcal{E})$; see
Fig.~\ref{fig:1}(c).  Hence, from the perspective of information
theory, this discrimination scheme is completely
non-destructive as $\bar{D}=0$ and $\bar{R}=1-\bar{E}$.

The implementation for schemes of the Kennedy type \cite{Kennedy73},
that unambiguously discriminate one of the signals, was also
investigated in Ref.~\cite{Han:2017b} using an atomic receiver.  In
order to unambiguously discriminate the state $\ket{\alpha}$, the
signal set is displaced by $D(\alpha)$ to $\{\ket{2\alpha},\ket{0}\}$.
If the atom, initially prepared in its ground state $\ket{\mathrm{g}}$, is
detected in the excited state $\ket{\mathrm{e}}$, the decision that the signal
state is $\ket{\alpha}$ can be made with certainty and no sequential
measurement is needed.  If the atom is detected in $\ket{\mathrm{g}}$, more
information can be extracted by subsequent measurements on the
post-measurement state (for $|\alpha|^2>0$).  However, in this scheme,
part of the information is destroyed due to the atomic measurement,
and the accessible information cannot be fully recovered through any
subsequent measurement, i.e.,
$I'_{\text{max}}(\mathcal{E},\Pi)<I_{\text{acc}}(\mathcal{E})$ as long as
$\bar{E}\neq0$ and $|\alpha|^2>0$; see Fig.~\ref{fig:1}(d). The
sequential measurement scheme for this unambiguous discrimination
strategy has also been investigated, and a significant increment in the
extracted information through subsequent measurements has been
demonstrated as shown in Fig.~\ref{fig:1}(d).

\textit{Discussion}.---
For the problem of quantum state discrimination, and in
particular quantum receivers, one aims at gaining maximal classical
information from the quantum state.  Our approach, based on mutual
information, is not only directly linked to the capacity of the
resulting classical communication channel, but allows moreover to
quantify how much additional information could be obtained by
subsequent measurements. It is the trade-off between the fraction of measured, residual and destroyed information that well characterizes the performance of a quantum measurement for state discrimination. The method can, for example, be used to
analyze sequential measurement schemes for any number of signal
states, or a multi-partite scenario with local measurements and classical communication of the measurement outcomes.

\end{document}